\documentclass[runningheads]{llncs}
\usepackage[T1]{fontenc}
\usepackage{graphicx}
\usepackage{amsmath}
\usepackage{tabularx}
\usepackage{enumitem}
\usepackage{caption}
\usepackage{subcaption}
\usepackage{xspace}
\usepackage{cite}
\usepackage{booktabs}
\usepackage{makecell,multirow,multicol}
\usepackage{xcolor}

\usepackage{hyperref,color}

\def\basegpt{GPT-5.4-mini\xspace}
\def\basesota{ShelLM\xspace}
\def\baseimpr{E-ShelLM\xspace}

\def\fullname{ShellGames\xspace}

\def\repo{\url{https://anonymous.4open.science/r/repo_sub_MTD-4874/}}

\title{ShellGames: Speculative LLM-Driven SSH Deception}

\begin{document}

\author{Umberto Salviati\inst{1,3}\orcidID{0009-0006-1475-9677} \and
Fabio De Gaspari\inst{2}\orcidID{0000-0001-9718-1044} \and
Mauro Conti\inst{1,4}\orcidID{0000-0002-3612-1934} \and
Luigi V. Mancini\inst{2}\orcidID{0000-0003-4859-219}}

\authorrunning{Salviati et al.}

\institute{Department of Mathematics, University of Padua, Padua, Italy\\
\email{umberto.salviati@phd.unipd.it, mauro.conti@unipd.it}\\
\and
Dipartimento di Informatica, Sapienza University of Rome, Rome, Italy\\
\email{\{degaspari,mancini\}@di.uniroma1.it}\\
\and Fondazione Bruno Kessler, Italy\\
\and Orebro Universitet, Örebro, Sweden}
\maketitle             

\begin{abstract}
Cyber deception and Moving Target Defense are promising strategies that aim to disrupt adversaries by increasing uncertainty. However, sustaining long-lived, credible interactive sessions with adversaries remains an open challenge. Large Language Models (LLMs) offer a promising path toward more dynamic deception systems, but suffer from key limitations that fundamentally limit their applicability, including: lack of persistent state, output inconsistencies, hallucinations, latency, and susceptibility to behavioral subversion that may reveal the deception.

We propose \fullname, an SSH shell simulator based on LLM designed to address these limitations. \fullname combines five complementary techniques: (i) Automatic Chain-of-Thought and few-shot learning to improve correctness; (ii) memory management to maintain system state coherency; (iii) speculative command execution to reduce response latency; (iv) smart routing of complex interactive commands to a sandboxed environment; and (v) subversion detection leveraging the constrained input-output domain of shell environments. To enable systematic evaluation, we introduce a standardized benchmarking protocol and dataset spanning correctness, consistency, state tracking, and robustness tasks. \fullname achieves $0.898$ command accuracy on correctness ($+5.3pp$ over baselines), $0.918$ sequence-level accuracy on consistency ($+36pp$), $0.98$ state tracking accuracy ($+18.3pp$), and $0.95$ accuracy on robustness ($+37pp$). A user study with $n=20$ participants confirms that \fullname achieves realism comparable to a real shell under free exploration and outperforms traditional honeypots on perceived command coverage.

\end{abstract}
\section{Introduction}

In recent years, cyber defense has been shifting from static, reactive mechanisms towards dynamic and proactive strategies aimed at disrupting and misleading adversaries. Cyber Deception~\cite{jajodia2016cyber} and Moving Target Defense (MTD)~\cite{jajodia2011moving} have emerged as complementary approaches that increase attacker uncertainty, imposing an asymmetric disadvantage by presenting a dynamic attack surface that is difficult to map and exploit~\cite{cho2020toward,lei2018moving}. Despite their promise, these approaches face significant challenges in providing credible, in-depth deception, particularly during long-lived, interactive sessions where coherence across interconnected services presented to the adversary must be maintained~\cite{pagnotta2023dolos}.

Recent advances in deep learning, and Large Language Models (LLMs) in particular, suggest a potential path toward more dynamic and agentic cyber defense systems~\cite{alam2024ctibench,kassianik2025llama,ren2025automated}. By enabling flexible, context-aware response generation, LLMs can potentially overcome many of the drawbacks of classical MTD and deception approaches, without requiring fully realized system deployments. Their capabilities in contextual reasoning, sustained interactive dialogue, and implicit knowledge of system configurations make them particularly attractive for deception scenarios. However, current applications of LLMs in security remain limited, particularly in highly interactive and adversarial settings~\cite{rahman2025survey,elzemity2025cyberllminstruct}. Notably, LLMs lack a grounded, persistent system state beyond their finite context window and system prompts, leading to temporal inconsistencies~\cite{hatalis2023memory}. Their probabilistic nature may lead to semantically incorrect or hallucinated outputs and failure to accurately model execution semantics~\cite{bang2025hallulens}. Furthermore, their inability to clearly separate instructions from input data exposes LLMs to adversarial manipulation, such as prompt-injection attacks~\cite{greshake2023not}. Finally, their computational overhead can introduce latency that may alert the adversary. 

These limitations point to a broader gap in current defenses: the absence of stateful, in-depth interactive environments capable of sustaining long-lived interactions with adversaries while maintaining coherent and realistic behavior. This gap motivates a shift toward what we term \textit{agentic cyber deception systems}; defense systems that maintain internal models of the environment, reason about adversary behavior, and dynamically adapt their actions to achieve objectives such as deception, adversary engagement, and intelligence gathering. As a first step in this direction, we propose \fullname, an LLM-driven SSH shell simulation that addresses key limitations of prior LLM-based approaches, including short memory, output inconsistency, hallucinations, and unrealistic response latency. \fullname provides a coherent, stateful interactive interface by combining several complementary techniques. It employs Automatic Chain of Thought~\cite{zhangautomatic} and few-shot learning~\cite{brown2020language} to minimize syntactic and semantic errors, and integrates ad-hoc memory management to limit inconsistencies and hallucinations. Inspired by CPU architecture, \fullname improves responsiveness through speculative command execution, which anticipates likely user commands and precomputes responses to reduce latency. Additionally, \fullname utilizes a smart routing mechanism to transparently offload complex interactive commands (e.g., \textit{nano}) to a sandboxed backend execution environment. Finally, \fullname incorporates a subversion detection mechanism that exploits the constrained input-output domain of shell environments to identify adversarial inputs and responses, defending against subversion attempts.
To systematically evaluate LLM-based shells, we introduce a standardized benchmarking protocol and a curated evaluation dataset. We measure performance across multiple consistency and robustness metrics, enabling a direct comparison of \fullname with existing state-of-the-art approaches and demonstrating the advantages of our design. Finally, we conduct a small-scale user study to assess \fullname's coherence and realism in a real-world setting. 

In summary, we make the following contributions:
\begin{itemize}
    \item We propose \fullname, an LLM-driven SSH shell simulation that addresses key limitations of current LLM-based approaches, including short memory, inconsistent outputs, hallucinations, unrealistic response timing, and susceptibility to subversion.
    \item We introduce several techniques that enhance realism and responsiveness, including Automatic Chain of Thought, few-shot learning, ad-hoc memory management, speculative command execution, subversion detection, and smart routing of complex commands.
    \item We design a standardized evaluation protocol and dataset\footnote{\repo} for LLM-based shells, enabling systematic assessment across multiple consistency and robustness metrics, and compare \fullname against state-of-the-art. 
    \item We conduct a small-scale user study to provide real-world validation of the system’s realism and effectiveness.
\end{itemize}

\section{Background and Related Work}

\paragraph{Honeypots.} 
Honeypots are proactive cybersecurity mechanisms that protect vulnerable systems by deceiving malicious actors. Their main goals are to gather threat intelligence, profile adversary behavior, and analyze new intrusion techniques \cite{ILG2023103737, 1254322}. They are typically classified by their level of interaction with adversaries: low-, medium-, and high-interaction systems \cite{10.1145/1233341.1233399}. Low-interaction honeypots simulate specific services at scale but offer limited adversary engagement \cite{1495930, 8717918}. Medium-interaction systems like Cowrie\cite{cowrie} offer realistic SSH/Telnet emulation in controlled environments. High-interaction honeypots run full operating systems to capture sophisticated attacks, providing high-fidelity environments but incurring significant resource costs, containment risks, and complexity~\cite{ILG2023103737}. Traditional honeypots are static, limiting their effectiveness against modern, adaptive threats.

\paragraph{Adaptive and Intelligent Honeypots}
Recent work addresses these limitations by using deep learning to build intelligent deception systems, from application-specific honeypots \cite{AIIpot, HSQL} to full-system emulation \cite{sladic2024llm,guan2024honeyllm, Honey_llama, llmHoneypot }. However, these efforts are fragmented, with little progress toward a unified framework or common baseline, aside from \cite{DontStop}. Major architectural and operational gaps persist: systems typically use bare LLMs without robust, stateful support, ignore practical constraints such as latency, and inadequately address core security risks like adversarial manipulation and prompt injection.
\section{Threat Model}
\label{sec:threat_model}

We consider a setting in which an adversary obtains access to a service that exposes an interactive command line interface, such as Secure Shell (SSH). The defender deploys \fullname as a deceptive shell environment, designed to emulate realistic interactive sessions. The \fullname system aims to sustain realistic, long-lived interactions, delay the adversary's progress, and provide valuable threat intelligence.
Crucially, \fullname never exposes the real underlying infrastructure. Our focus is on post-compromise interaction with \fullname; we do not address intrusion prevention or authentication mechanisms, but rather the behavior of \fullname after adversarial access has been established.

\paragraph{\textbf{Adversary Model.}}
We consider an adversary that can issue arbitrary shell commands through a remote command-line interface over long-lasting interactive sessions. The adversary may attempt to probe \fullname or manipulate it through adversarial inputs, including prompt injection or commands crafted to induce inconsistencies or errors. We assume that the adversary is not aware of the presence of \fullname, but we consider a strong adversary that may adapt their strategy based on the responses it observes. The adversary’s interaction is restricted to the exposed shell interface and does not include any out-of-band visibility into the underlying infrastructure.

\paragraph{\textbf{Defender Goals.}}
The defender aims to provide a coherent, stateful, and credible shell environment that simulates a real command line interface (e.g. Bash). The shell environment should minimize artifacts that could reveal the deceptive nature of the system, as well as support the collection of interaction data with the adversary for intelligence analysis.
\section{System Design}
\label{sec:methodology}

Existing LLM-based interactive environments present several limitations. They typically lack persistent state, resulting in inconsistencies over multi-step interactions. Their probabilistic generative nature may lead to incorrect or hallucinated responses, violating expected shell semantics. They also struggle to handle complex or interactive commands and introduce persistent latency that can undermine realism. These shortcomings are exacerbated by prolonged, multi-step sessions where adversaries probe system behavior.

\fullname is a speculative, LLM-driven SSH deception system designed to address these challenges and provide a stateful, coherent and responsive shell environment. To this end, \fullname integrates five complementary components: i) Automatic Chain of Thought~\cite{zhangautomatic} and few-shot learning~\cite{brown2020language} to minimize syntactic and semantic errors; (ii) memory management to limit inconsistencies and hallucinations over sustained interactions; (iii) speculative command execution, inspired by CPU speculative execution, to anticipate likely user commands and precompute responses to reduce latency; (iv) smart routing, which offloads complex interactive commands (e.g., nano) to a backend execution environment; and (v) subversion detection, which detects subversion attempts and returns a shell-compliant error response. By combining these components, \fullname provides a consistent, responsive, and credible environment while laying the groundwork toward agentic cyber deception systems. We present the architecture of \fullname in Section~\ref{sec:architecture} and defer implementation details to Appendix~\ref{sec:impl_details}.

\begin{figure}[t]
    \centering
    \includegraphics[width=\linewidth]{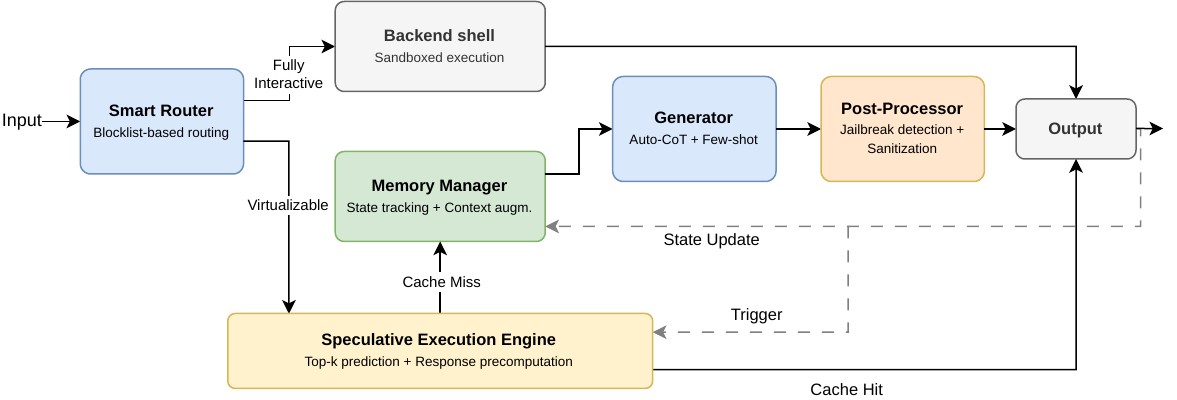}
    \caption{\fullname high-level overview.}
    \label{fig:architecture}
\end{figure}

\subsection{Architecture Overview}
\label{sec:architecture}
Figure~\ref{fig:architecture} provides a high-level overview of \fullname architecture and its main components. \fullname is designed as a hybrid system that orchestrates multiple components in a simulated stateful shell environment. At a high level, the system processes inputs through a pipeline that combines system state management (\textit{Memory Manager}), execution routing (\textit{Smart Router}), LLM-based generation (\textit{Generator}), command prediction (\textit{Speculative Execution Engine}), and subversion detection (\textit{Post-Processor}). 

Upon receiving an input command, the Smart Router component analyzes it and directs the execution along one of two paths: commands that require full interactivity are delegated to the sandboxed backend execution environment, while all other commands are forwarded to the Memory Manager. The Memory Manager constructs an augmented prompt that combines the input command, the initial system definition, and relevant contextual state, then passes it to the Generator. The Generator produces a response, which is then evaluated by the Post-Processor to detect and sanitize potential subversion attempts. In parallel, the Memory Manager processes the output to update its internal state, while the Speculative Execution Engine precomputes responses for a set of likely follow-up commands. This design cleanly separates LLM-based generation, state management, complex command execution, and behavioral prediction, with each component targeting distinct limitations of prior LLM-based approaches.

\paragraph{\textbf{Generator.}}
The Generator is the core response synthesis module of \fullname. It leverages an LLM to generate command responses that are coherent with the system definition---an initial system prompt that encodes general behavioral guidelines and the simulated host configuration---and the relevant system state provided by the Memory Manager. Formally, this context acts as a prior that conditions the model's output distribution toward responses that are consistent with prior interactions and expected shell semantics. 

To reduce syntactic and semantic errors, the Generator leverages Automatic Chain of Thought (Auto CoT) prompting~\cite{zhangautomatic} to induce the model to produce explicit, intermediate steps before providing a final response. Furthermore, few-shot learning~\cite{brown2020language} is used to condition generation on representative examples of expected shell syntax and semantics. This structured generation process ensures that outputs are grounded in the current system state and consistent with expected shell behavior, improving syntactic and semantic correctness.

\subsubsection{Memory Manager}
To address the lack of persistent state in LLMs, \fullname incorporates an explicit memory management module that maintains a structured representation of the simulated system state. The Memory Manager serves two primary functions: \textit{context augmentation}, which retrieves relevant state information and includes it in the Generator's prompt; and \textit{state update}, which updates the internal system state representation based on simulated commands. The system state comprises a pruned command history and a virtual filesystem data structure. Both structures are asynchronously updated by a background synthesis job that lazily processes recent interactions, separating state maintenance from the critical response generation functions. The synthesis job carries out two main tasks: (i) \textit{virtual filesystem update}, which parses recent state-changing commands and updates an internal tree structure representing the filesystem hierarchy and contents; and (ii) \textit{command history pruning}, which removes non-state-changing commands and filesystem operations already captured by the virtual filesystem, keeping the Generator's context concise.

The final Generator's context comprises the pruned command history, the virtual filesystem, and the current command. By externalizing system state into dedicated data structures, the Memory Manager reduces output inconsistencies and enables coherent, stateful interactions over long-lasting sessions, while the compact representation reduces inference latency.

\subsubsection{Speculative Execution Engine}
Inspired by speculative execution in CPUs, the Speculative Execution Engine mitigates the inherent latency of LLM inference. Based on a compact representation of the recent command history, a lightweight prediction model trained on shell interaction traces produces a ranked set of top-$k$ likely follow-up commands. Candidate commands are evaluated in parallel by multiple Generator instances, and responses are cached for potential future use. 

Upon receiving a new command, \fullname first queries the speculative cache. On a cache hit, the precomputed response is returned immediately, bypassing the Generator. On a cache miss, \fullname falls back to the standard generation pipeline. Incorrect predictions are silently discarded without affecting the system state, ensuring speculative execution does not introduce inconsistencies. 
Any state-changing operation invalidates the cache and triggers a new speculative execution pass, ensuring cached responses remain consistent with the current state.

\subsubsection{Smart Router}
Certain classes of commands are challenging to accurately simulate using an LLM alone, particularly those involving interactive interfaces (e.g., \textit{nano}) or runtime-dependent execution semantics (e.g., \textit{wget}). 
Simulating such commands purely through an LLM would lead to inconsistencies, broken interaction flows, and detectable artifacts that could reveal the nature of the system.

\fullname addresses this limitation through a smart routing mechanism that detects and delegates complex commands to a sandboxed backend shell environment. Routing decisions are based on a blocklist of commands known to require interactivity or runtime-dependent execution semantics. The backend environment is isolated for safety and transparently synchronized with the system state maintained by the Memory Manager, ensuring coherence between simulated and executed interactions. Relevant side effects, such as filesystem modifications, are reflected back in the internal system state maintained by the Memory Manager. This hybrid execution approach balances realism, robustness, and safety, while also minimizing detectable inconsistencies.

\subsubsection{Post-processor} 
A key vulnerability of LLM-based systems is their susceptibility to prompt-injection and jailbreaking attacks: adversarial inputs that exploit the model's inability to separate input data from instructions, with the goal of overriding intended behavior and eliciting out-of-character responses~\cite{greshake2023not,perezignore}. In general-purpose LLMs, detecting prompt-injection attacks is inherently difficult, as both the input and output spaces are essentially unbounded natural language. However, an SSH shell simulator operates over a much more constrained domain. Valid inputs are shell commands following a well-defined syntax. Outputs are structured responses with a defined format and content, bounded by shell semantics. These structural constraints make both 

\fullname leverages this property through a dedicated transformer-based classifier that evaluates both the input command and the generated response to detect subversion attempts. On the input side, the classifier identifies commands that deviate from valid shell syntax or exhibit semantics characteristic of prompt injection and jailbreaking. On the output side, the model identifies deviations from expected shell output structure and inconsistencies with the issued command. The detector uses in-context training with few-shot learning~\cite{brown2020language} to improve performance and limit overhead. Upon detecting an attack, \fullname returns a syntactically valid shell error message, thereby preserving the realism of the simulation.

\section{Evaluation Protocol and Dataset}
\label{sec:eval_protocol}
A key contribution of this work is the design of a standardized evaluation protocol for LLM-based shell simulators and a curated testing dataset of shell commands. The evaluation protocol systematically evaluates simulated shells along four key dimensions: 1) \textit{correctness}, measuring whether individual command responses are accurate; 2) \textit{consistency}, measuring whether the system maintains a coherent system state across multi-turn interactions; 3) \textit{filesystem state tracking}, as a targeted stress-test of dynamic state tracking and memory; and 4) \textit{robustness} against adversarial inputs such as jailbreaking~\cite{andriushchenkojailbreaking}. The following sections describe each evaluation dimension in detail.

\paragraph{Correctness Evaluation}
Command response correctness is evaluated through single-turn interactions, where each test instance consists of an isolated shell command paired with a reference response. Commands are sampled from a realistic distribution of human-generated shell commands, covering a representative mix of several types of operations. Since commands are evaluated independently, this benchmark focuses on baseline syntactic and semantic correctness in the absence of interaction history, isolating response quality from state tracking.

\paragraph{Consistency Evaluation}
Consistency evaluation assesses the capability of the system to correctly track interaction history across multi-turn sessions of 2--10 commands. All command sequences are generated following datasets of real execution traces. Consistency performance is evaluated both at the \textit{sequence} and \textit{command} level. At the sequence level, a session is considered correct only if all commands in the sequence are answered correctly. At the command level, each response is evaluated individually within its sequence context. Unlike the single-turn correctness evaluation, this setting requires the system to account for prior interaction history, rather than generating responses in isolation. This benchmark focuses on the system's ability to condition responses based on recent interaction history.

\paragraph{Filesystem State Tracking}
The filesystem state tracking evaluation stress-tests the system's ability to track state changes across long-lasting sessions. Sessions are constructed from synthetic traces comprising representative filesystem operations, such as file and directory creation/deletion, directory traversal, content output, and directory listing. The traces are created with multiple lengths comprising 50, 100, and 200 command sequences. We evaluate filesystem state tracking across two path-resolution settings. Absolute path sessions serve as a baseline, as commands are self-contained and do not require tracking of the current working directory. Relative path sessions are considerably more challenging, as the system must keep track of working directory changes across directory traversals and filesystem changes. Evaluating simulation systems under both settings enables us to isolate the complexity of directory-traversal state tracking and its impact on simulation fidelity.

\paragraph{Robustness Evaluation}
System robustness is evaluated through single-turn interactions with adversarially crafted inputs. Adversarial inputs are designed to destabilize or manipulate the system through jailbreaks, malformed or erroneous inputs, and prompt injection. This benchmark focuses on the system's ability to maintain coherent behavior in adversarial settings, a critical property for deployment in the real world.

\subsubsection{Dataset Details}

To support our evaluation protocol, we created a curated dataset of command-response shell interactions spanning all evaluation dimensions. The dataset comprises commands sampled from real-world logs and augmented with varying arguments, synthetic sequences designed to test specific simulation capabilities, and hand-crafted adversarial inputs targeting robustness and susceptibility to manipulation. To mimic the natural interaction sequence of humans, these commands were augmented using automated LLM-based tools to generate multiple variants. All commands were validated against a real shell, which was also used to obtain the corresponding ground-truth responses.

Table~\ref{tab:dataset} summarizes the dataset and the evaluation protocol.

\begin{table}[t]
\setlength{\tabcolsep}{3.2pt}
\centering
\scriptsize
\caption{Overview of the proposed evaluation protocol and dataset.}
\label{tab:dataset}
\begin{tabular}{cccccc}
\toprule
\makecell{\textbf{Task}} & 
\makecell{\textbf{Interaction}\\\textbf{Type}} & 
\makecell{\textbf{Sequence}\\\textbf{Length}} & 
\makecell{\textbf{Total}\\\textbf{Commands}} & 
\makecell{\textbf{Data}\\\textbf{Source}} & 
\makecell{\textbf{Evaluation}\\\textbf{Goal}} \\
\midrule
Correctness & Single-turn   & 1         & 197   & Augm. real-world    & Response accuracy \\
Consistency & Multi-turn    & 2--10     & 1396  & Augm. real-world    & Interactive consistency \\
Filesystem  & Multi-turn    & 50--200   & 350  & Synthetic     & FS state tracking  \\
Robustness  & Single-turn   & 1         & 40    & Adversarial   & Adversarial robustness \\
\bottomrule
\end{tabular}
\end{table}

\section{Experimental Analysis}
\label{sec:evaluation}

\subsection{Baselines and Metrics}

We select three baseline approaches for our comparison: \basegpt, \basesota, and \baseimpr. \basegpt represents the current state-of-the-art in general-purpose LLMs and was selected to provide an off-the-shelf baseline. It belongs to a new generation of ``reasoning'' models that explicitly allocate a budget of tokens for internal reasoning before providing a response. We explicitly prompt the model to simulate a shell interface (see Appendix~\ref{sec:impl_details} for the detailed formulation). \basesota~\cite{sladic2024llm} represents the current open-source state-of-the-art LLM shell simulator. It leverages CoT and purpose-built prompt engineering to provide a realistic shell environment. For a fair comparison, we update the original implementation to use GPT-4.1-mini, the same model used in \fullname (see Section~\ref{sec:impl_details}). \baseimpr is an enhanced variant of \basesota that we introduce as an additional baseline. We add an enhanced prompting strategy and a more explicit context prior, providing a more coherent initial system state. 

\paragraph{Performance Metrics.}
We assess system performance through two complementary metrics: command-level accuracy and sequence-level accuracy. Command-level accuracy evaluates the correctness of an individual command-response pair in isolation. It is computed as the ratio of correct responses to the total number of commands issued. Sequence-level accuracy evaluates the correctness of an entire sequence of $N$ commands, where each individual response must be correct for the full sequence to be considered correct. It is computed as the ratio of fully correct sequences of responses to the total number of sequences. We evaluate response correctness using a strict \textit{exact string matching} criterion, where responses are matched to a ground truth produced by a real shell environment.

\subsection{Evaluation Protocol Results}
We compare \fullname to all baseline approaches under the evaluation protocol described in Section~\ref{sec:eval_protocol}, covering command correctness, consistency, filesystem state tracking, and robustness.

\subsubsection{Correctness}
We evaluate correctness under single-turn interaction on 197 total commands. We partition the commands into four categories: mathematical operations, programming, text processing, and miscellaneous. Table~\ref{tab:correctness} reports our results. \fullname achieves the highest overall accuracy, $0.898$, matching or outperforming baselines in all categories. All models perform well on programming tasks. On Text Processing and miscellaneous, \fullname\ consistently achieves the highest accuracies ($\sim0.935$), with significant gains over \basesota and \baseimpr specifically in the miscellaneous category. This improvement is likely attributable to the specialized Auto CoT strategy of \fullname's Generator, which enhances the model's ability to handle complex transformations and miscellaneous tasks. The Math category is particularly challenging, as it features complex operations such as hashing and masking. \fullname outperforms all baselines in this category, with an $11.4$ percentage points (\textit{pp}) improvement over \basegpt and $\sim3pp$ over \basesota. An error analysis reveals that the majority of mistakes produced by \fullname are not catastrophic failures, but rather plausible approximations of the expected response. Discrepancies frequently arise in tasks involving hashing: the model produces syntactically valid outputs that do not match the true hash of the input. Because our evaluation enforces exact-string matching, such deviations are penalized as errors, despite their plausibility to a human observer.

\begin{table}[t]
\centering
\scriptsize
\caption{Correctness comparison across different classes of data. Command-level accuracy reported.
}
\label{tab:correctness}
\begin{tabular}{lcccc}
\toprule
\textbf{Task} & \textbf{\basegpt} & \textbf{\basesota} & \textbf{\baseimpr} & \textbf{\fullname (Ours)} \\
\midrule

Math            & 0.700 & 0.786 & 0.771 & \textbf{0.814} \\
Programming     & 0.938 & 0.938 & 0.938 & \textbf{1.000} \\
Text Processing & 0.886 & 0.911 & 0.902 & \textbf{0.935} \\
Miscellaneous   & 0.875	& 0.625	& 0.688	& \textbf{0.938} \\
\midrule
\textbf{Total Accuracy} & 0.822 & 0.848 & 0.843 & \textbf{0.898} \\
\bottomrule
\end{tabular}
\end{table}

\subsubsection{Consistency}
\begin{figure}[t]
\centering
\begin{subfigure}{0.495\columnwidth}
    \includegraphics[width=\textwidth]{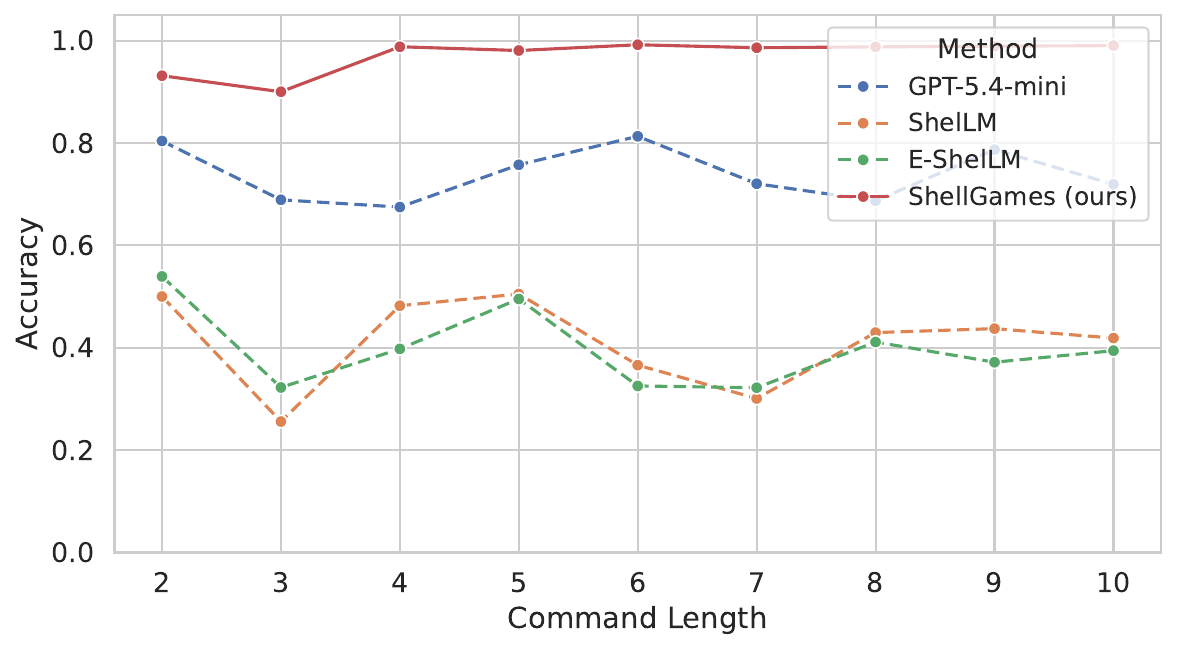}
    \caption{Command-level accuracy.}
    \label{fig:cmd-acc}
\end{subfigure}
\hfill
\begin{subfigure}{0.495\columnwidth}
    \includegraphics[width=\textwidth]{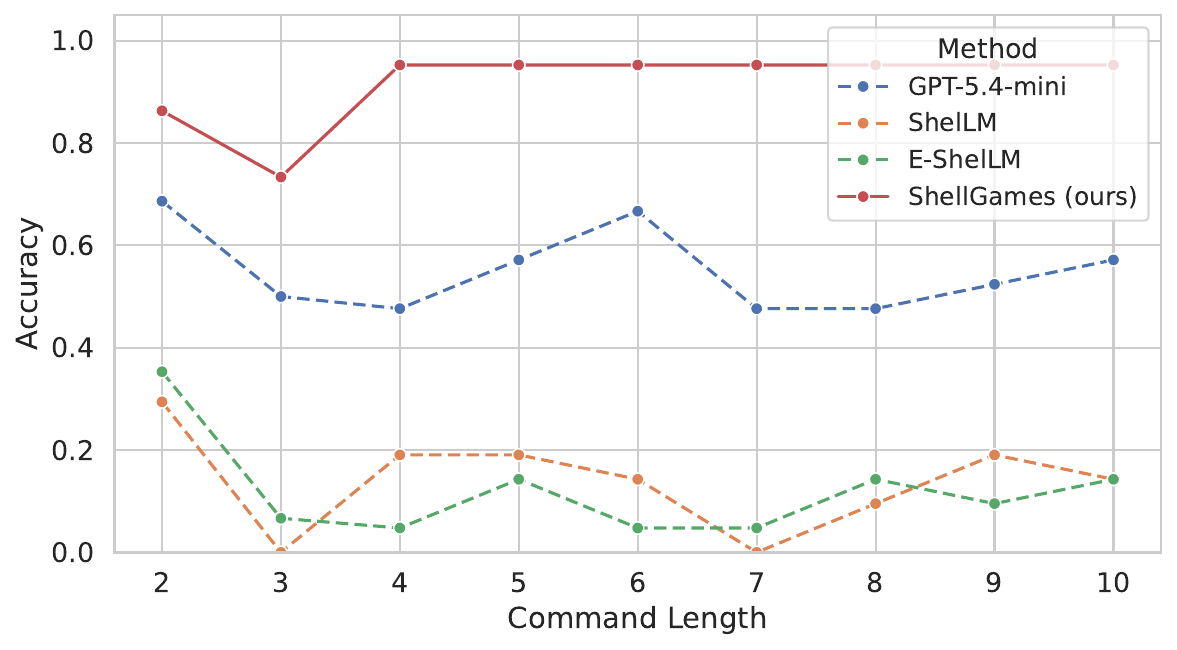}
    \caption{Sequence-level accuracy.}
    \label{fig:seq-acc}
\end{subfigure}
\caption{Consistency comparison across varying sequence lengths.}
\end{figure}

We evaluate consistency at both the sequence and command levels. At the sequence level, a session is considered correct if all commands in the sequence are answered correctly. At the command level, each response is evaluated individually within its sequence context. Figures~\ref{fig:cmd-acc} and~\ref{fig:seq-acc} illustrate command- and sequence-level accuracy across interaction sequences of length 2--10, respectively.

\paragraph{Command-level consistency.}
\fullname achieves a mean command-level accuracy of $0.971$, compared to $0.739$ for \basegpt and approximately $0.4$ for both \basesota and \baseimpr. A comparison between command-level consistency and correctness performance (Table~\ref{tab:correctness}) provides interesting insights. While command-level performance is evaluated independently for each command, as with correctness, the model must condition its output on an evolving interaction history. Therefore, command-level consistency additionally captures the ability to maintain a coherent system state across turns. The substantial drop in command-level consistency observed for \basesota and \baseimpr (approx. $\-43pp$) confirms that their performance degrades significantly when state tracking is required. Similarly, \basegpt shows a moderate $-7pp$ performance delta between correctness and command-level consistency.
By contrast, \fullname's performance exceeds its single-turn correctness ($+7.76$pp), indicating that the Memory Manager actively improves response quality by providing key contextual information.

\paragraph{Sequence-level consistency.}
\fullname achieves a mean sequence-level accuracy of $0.918$, outperforming the closest baseline (\basegpt) by over $36$ percentage points (pp; $0.549$). \basesota and \baseimpr perform poorly under this metric, with mean accuracies of $0.138$ and $0.120$, respectively. These results reflect two fundamental limitations of context-window-based approaches. First, the finite context windows cause the model to lose track of earlier interactions as session length increases, leading to errors in responses that depend on prior state. Second, LLM performance degrades as relevant information moves to the middle of the context window~\cite{liu2024lost}, leading to increased errors. Both effects compound in multi-turn shell sessions, leading to substantially degraded performance. \fullname reaches $\sim0.95$ accuracy and maintains this level throughout sequence length 10, indicating strong contextual consistency within this range. A decrease in performance is observed at length 3 across all models, likely suggesting an artifact of the synthetic evaluation sequence rather than model-specific behavior.

\paragraph{Consistency Gap.}
The gap between sequence- and command-level performance highlights the complexity of maintaining state consistency across multi-turn interactions and, ultimately, the difficulty of presenting a credible system to adversaries. We term this error \textit{consistency gap}. \fullname shows a mean consistency gap of $5.35pp$ across all sequence lengths, highlighting the effectiveness of our techniques in ensuring system coherence. The comparatively larger consistency gap for \basegpt ($18.92pp$) indicates that its errors tend to be distributed across sequences, rather than cascading. This suggests syntactic or semantic errors, rather than hallucinated outputs that would lead to cascading mistakes.

\begin{table}[t]
\centering
\scriptsize
\setlength{\tabcolsep}{4pt}
\caption{State-tracking comparison across different sequence lengths and path-resolution settings. Command-level accuracy reported.}
\label{tab:fs_tracking}
\begin{tabular}{lccccc}
\toprule
\textbf{Path Type} & \textbf{Length} & \textbf{\basegpt} & \textbf{\basesota} & \textbf{\baseimpr} & \textbf{\fullname} \\
\midrule
\multirow{3}{*}{Relative} 
& 50  & 0.960  & 0.700 & 0.700 & \textbf{0.980} \\
& 100 & 0.960  & 0.600 & 0.600 & \textbf{0.990} \\
& 200 & 0.840  & 0.170 & 0.300  & \textbf{0.985} \\  \cmidrule{2-6}
& Mean& 0.920     & 0.490    & 0.533 & \textbf{0.985} \\ \midrule
\multirow{3}{*}{Absolute}
& 50  & \textbf{1.0}     & 0.380 & 0.740 & 0.960 \\
& 100 & 0.270  & 0.260 & 0.600  & \textbf{0.980} \\
& 200 & 0.750  & 0.370 & 0.370 & \textbf{0.990} \\ \cmidrule{2-6}
& Mean& 0.673  & 0.336 & 0.570    & \textbf{0.976} \\
\bottomrule
\end{tabular}
\end{table}

\subsubsection{Filesystem State Tracking} 
Table~\ref{tab:fs_tracking} reports command-level accuracy across filesystem interaction sequences of lengths 50, 100, and 200 under relative and absolute path-resolution settings. \fullname achieves the best mean accuracy in both settings, demonstrating robust long-horizon filesystem state tracking. In the relative path setting, \fullname consistently outperforms all other baseline approaches across all sequence lengths, showing no meaningful degradation as command history grows. \basegpt achieves competitive performance on shorter sequences ($0.96$ accuracy), but its performance degrades on the 200-command sequence ($0.84$, $-14.5pp$ compared to \fullname). This result suggests that \basegpt internal reasoning can help compensate for the absence of explicit memory management over short horizons. However, as the context size increases, it becomes insufficient and the model loses track of the system's state. \basesota and \baseimpr perform discreetly on short sequence lengths, reaching $0.7$ accuracy. However, their performance drops rapidly as sequence length increases, underscoring the importance of dedicated state-tracking functions.

The absolute path setting removes the need to dynamically track the working directory and should provide a simpler baseline for stateless approaches. Surprisingly, \basegpt and \basesota underperform in this setting compared to their performance on relative paths. \baseimpr mean accuracy increases by $3.7pp$, reflecting improved performance on long-horizon tracking ($+7pp$). \fullname achieves consistently high performance across all sequence lengths, outperforming the closest baseline approach by $30pp$.

\subsubsection{Robustness Evaluation.}
The robustness benchmark comprises 40 adversarial inputs spanning direct prompt injection, indirect prompt injection, and malformed commands. In direct prompt injection, the adversary explicitly crafts inputs to override system instructions and manipulate the model's behavior. Indirect prompt injection employs more covert manipulation vectors to achieve the same objective~\cite{greshake2023not}. Malformed commands introduce syntactic noise that can cause models to generate unexpected results. Table~\ref{tab:robustness_results} reports command-level accuracy across the three adversarial input categories for all approaches. 

\fullname achieves the highest overall accuracy, substantially outperforming the best baseline (\basesota) by $37.5pp$. The most pronounced gap is observed on direct prompt injection, where \basegpt completely fails, and \basesota and \baseimpr achieve only $0.188$ accuracy. Robustness to indirect prompt injection is similarly low across the baselines, with the sole exception of \baseimpr. This vulnerability is expected, as none of the baseline approaches implements any explicit defense mechanism against adversarial inputs. \basegpt's complete failure is particularly notable, as its reasoning capabilities do not appear to provide any advantage against adversarial inputs. 

\fullname effectively neutralizes both injection attack classes, achieving $0.938$ accuracy on direct prompt injection and perfect accuracy on indirect prompt injection. As we demonstrate in our ablation study (see Section~\ref{sec:ablation}), this is primarily due to the Post-processor module's jailbreak detector, which identifies adversarial inputs before providing the final response. All approaches perform reasonably well on malformed inputs. However, our proposal still achieves the highest accuracy, showing robustness to both deliberate adversarial manipulations and unintended input noise.

\begin{table}[t]
\centering
\scriptsize
\caption{Robustness accuracy across multiple tasks. Command-level accuracy.}
\label{tab:robustness_results}
\begin{tabular}{lcccc}
\toprule
\textbf{Category} & \textbf{GPT-5.4 Mini} & \textbf{\basesota} & \textbf{\baseimpr} & \textbf{\fullname (Ours)} \\
\midrule
Indirect Prompt Injection & 0.500 & 0.500 & 0.750 & \textbf{1.000} \\
Direct Prompt Injection & 0.000 & 0.188 & 0.188 & \textbf{0.938} \\
Malformed Command & 0.850 & 0.900 & 0.800 & \textbf{0.950} \\
\midrule
\textbf{Total Accuracy} & 0.475 & 0.575 & 0.550 & \textbf{0.950} \\
\bottomrule
\end{tabular}
\end{table}

\subsection{Ablation Study}
\label{sec:ablation}

\paragraph{Memory Manager}
We quantify the impact of the Memory Manager (MM) on state tracking with an ablation study on the filesystem state tracking task. Table~\ref{tab:filesystem_ablation} presents our results. The MM provide consistent accuracy increases for both relative and absolute path settings. For the relative paths, MM provides significant consistency improvements on long sequences, where the default context window struggles to maintain the state. Interestingly, the largest mean gain is obtained in the absolute path settings. This is likely thanks to the explicit virtual filesystem structure that enables easier direct matching of the absolute path to the tree structure.

\begin{table}[t]
\setlength{\tabcolsep}{6pt}
\centering
\scriptsize
\caption{Module Manager (MM) ablation study on state tracking performance.}
\label{tab:filesystem_ablation}
\begin{tabular}{lcccc}
\toprule
\textbf{Path Type} & \textbf{Length} & \makecell{\textbf{\fullname}\\w/out MM} & \makecell{\textbf{\fullname}\\with MM} & \textbf{MM $\Delta$ }\\
\midrule
\multirow{4}{*}{Relative} 
& 50  & \textbf{1.0} & 0.980 & -0.02  \\
& 100 & \textbf{1.0} & 0.990 & -0.01  \\
& 200 & 0.865  & \textbf{0.985} & +0.12 \\ \cmidrule{2-5}
& Mean& 0.955 & \textbf{0.985} & +0.03 \\ \midrule
\multirow{4}{*}{Absolute} 
& 50  & 0.740  & \textbf{0.960} & +0.22 \\
& 100 & 0.850  & \textbf{0.980} & +0.13 \\
& 200 & \textbf{0.995}  & 0.990 & -0.005\\ \cmidrule{2-5}
& Mean& 0.861 & \textbf{0.976} & +0.115 \\
\bottomrule
\end{tabular}
\end{table}

\begin{table}[t]
\setlength{\tabcolsep}{5pt}
\centering
\scriptsize
\caption{Post-processor (PP) ablation study on robustness performance.}
\label{tab:robustness_ablation}
\begin{tabular}{lccc}
\toprule
\textbf{Category} & \makecell{\textbf{\fullname}\\w/out PP} & \makecell{\textbf{\fullname}\\with PP} & PP $\Delta$ \\
\midrule
Indirect Injection & 0.750          & \textbf{1.000} & +0.25  \\
Prompt Injection   & 0.000          & \textbf{0.938} & +0.938 \\
Malformed Command  & \textbf{0.950} & \textbf{0.950} & +0.0   \\
\bottomrule
\end{tabular}
\end{table}

\paragraph{Post-processor}
We quantify the impact of the Post-processor module on adversarial inputs with an ablation study on the robustness task. Results in table~\ref{tab:robustness_ablation} demonstrate that the Post-processor is essential for mitigating adversarial manipulations. Without the Post-processor, the base model fails catastrophically on prompt injections (0.0) and shows substantial weakness against indirect injections (0.750). The Post-processor dramatically improves robustness, increasing response accuracy on direct injections by $93.8pp$ and on indirect injections by $25pp$. Finally, the Post-processor does not degrade performance on benign inputs, as shown in Table~\ref{tab:detector_performance} in the Appendix.

\begin{figure}[t]
    \centering
    \includegraphics[width=0.8\columnwidth]{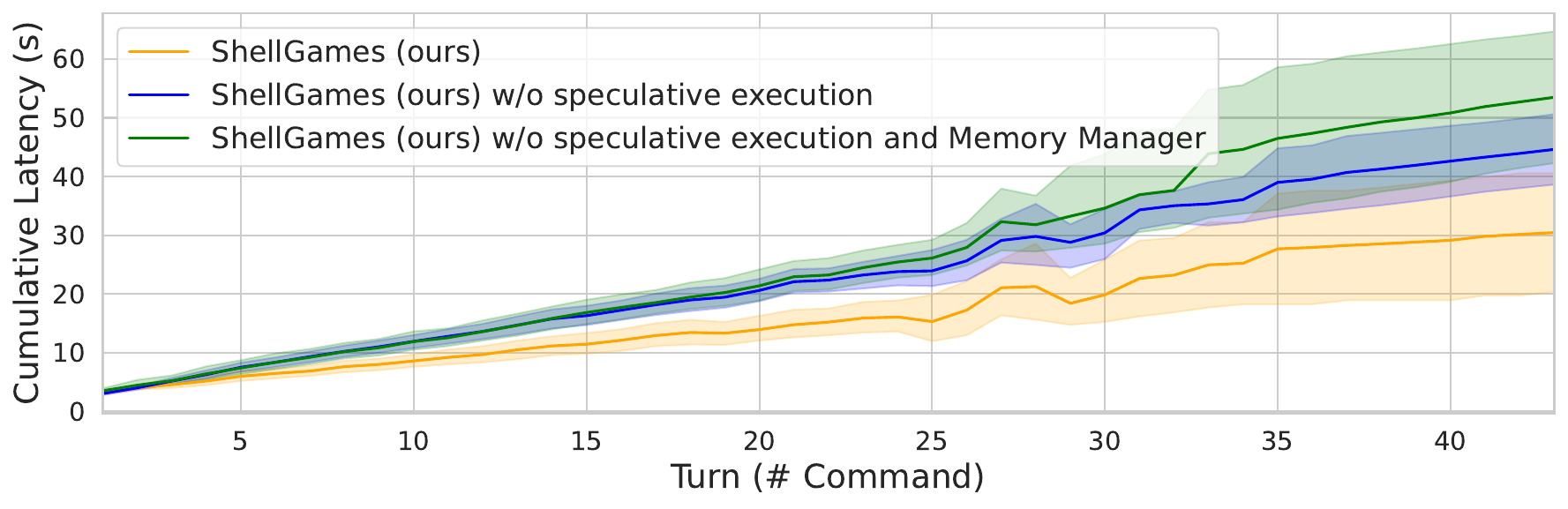}
    \caption{Cumulative response latency over command sequences of varying length.}
    \label{fig:time_ablation}
\end{figure}

\paragraph{Speculative Execution Engine}
We assess the impact of the Speculative Execution Engine (SEE) on system latency via ablation. We use real command traces collected during the user study for the evaluation. Notably, these traces are completely disjoint from the training data of the speculative predictor, providing a strict out-of-distribution evaluation of generalization ability. Figure~\ref{fig:time_ablation} illustrates cumulative latency over command sequences of varying length. Without SEE or memory management, \fullname exhibits a mean response time of $1.09\pm1.27s$. Introducing the Memory Manager yields a modest latency reduction of 8.82\%. The full architecture combining the Memory Manager and SEE reduces mean response time to $0.71\pm1.05s$, leading to a substantial overall latency decrease of 34.84\%. These empirical results confirm that our lightweight speculative approach substantially reduces latency without compromising functional accuracy.

\subsection{User Study}
\begin{figure}[t]
    \centering
    \includegraphics[width=\columnwidth]{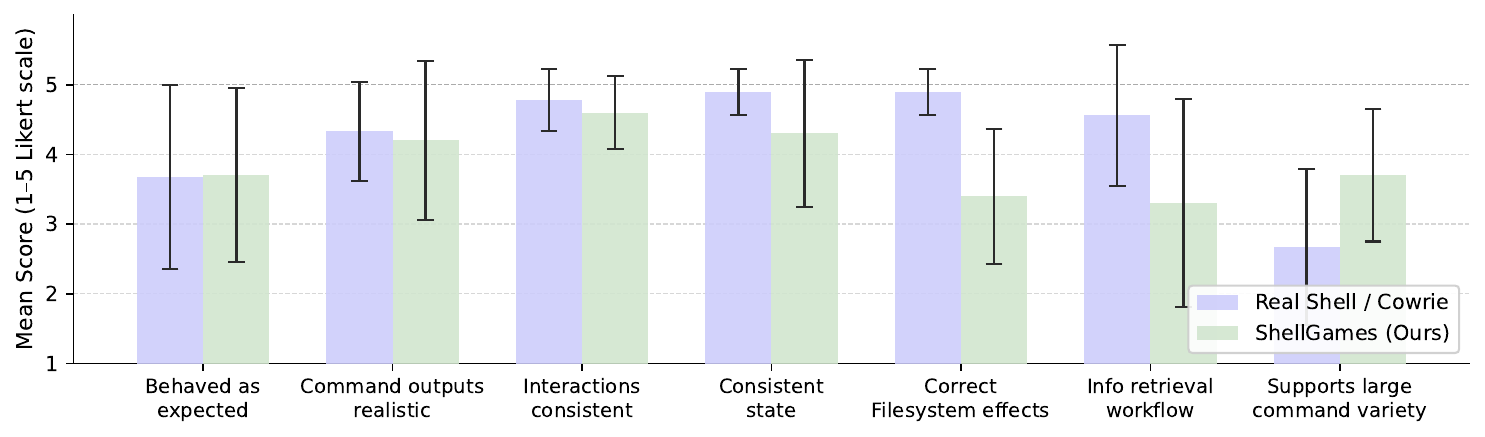}
    \caption{User study responses on a 5-point Likert scale.}
    \label{fig:user_study}
\end{figure}

We assess the perceived realism, interactive usability, and deceptive effectiveness of \fullname through an exploratory user study with $n=20$ voluntary participants. The study is structured around three tasks, each targeting different aspects of shell simulation fidelity. Participants were randomly assigned to either a real or \fullname-simulated environment for each task. Real environments comprised either a native bash shell (Tasks 1,2) or Cowrie~\cite{cowrie} in shell mode, an industry-standard medium-interaction honeypot (Task 3). Following each task phase, participants completed a standardized questionnaire evaluating perceived realism, state consistency, filesystem fidelity, and overall command coverage and capability based on a 5-point Likert scale. Figure~\ref{fig:user_study} presents our results.

\paragraph{Task 1: Unconstrained Interaction.}
Participants freely explored either a real shell or \fullname, providing a baseline assessment of perceived realism under naturalistic conditions. The primary aim was to determine whether participants could detect inconsistencies or artifacts indicative of a simulated environment. \fullname achieves scores comparable to the real shell baseline across all three questions in this task: expected behavior ($3.70$ vs. $3.67$), output realism ($4.20$ vs. $4.33$), and interaction consistency ($4.60$ vs. $4.78$). The negligible gap across all three metrics suggests \fullname is indistinguishable from a real shell under free exploration.

\paragraph{Task 2: Goal-Oriented Interaction.}
Participants were tasked with navigating and extracting information from a file system mirroring a well-known academic GitHub repository\footnote{https://github.com/academicpages/academicpages.github.io}. This task focuses on stress-testing consistency and filesystem fidelity. As expected, the real shell baseline performs well across state consistency ($4.89$) and filesystem correctness ($4.89$). \fullname performs well on state consistency ($4.3$), but shows a more significant gap on filesystem correctness ($3.4$) and information retrieval ($3.3)$. The large variance for \fullname on these metrics suggests inconsistent experience across participants, highlighting the complexity of ensuring filesystem fidelity.

\paragraph{Task 3: Comparative Deception against Traditional Honeypots.}
To encourage command exploration, in the final task, participants were asked to determine whether the system was a honeypot, directly comparing \fullname to Cowrie. A critical metric in this task is the perceived breadth of realistically supported commands, as traditional honeypots are constrained by rigid, predefined command sets. \fullname substantially outperforms Cowrie in this metric ($3.7$ vs $2.67$), suggesting that its LLM-based generation process supports meaningfully broader realistic command coverage. 

\paragraph{Discussion.} \fullname achieves a mean score of $3.89$, compared to $4.25$ for the real shell and Cowrie baseline. The gap is mainly driven by filesystem simulation inaccuracies, which become more apparent when users are primed to scrutinize filesystem-related command outputs. Under free exploration by unprimed users, \fullname achieves good realism, and outperforms traditional honeypots on perceived command coverage when users actively probe system capabilities. Together, these results validate the core objectives of \fullname's design and identify filesystem simulation as the main direction for future improvements.

\section{Conclusion}
\label{sec:conclusion}
We introduced \fullname, an LLM-based SSH deception system that overcomes the rigidity of traditional honeypots and the limitations of existing LLM approaches. By integrating memory management, speculative command execution, smart routing, and subversion detection, \fullname successfully balances high-fidelity shell simulation with operational efficiency and robust security.

Our evaluation demonstrated that \fullname achieves state-of-the-art performance across correctness, consistency, state tracking, and robustness. Our ablation analysis results provide direct validation for our architectural design. Finally, an exploratory user study confirmed the effectiveness of our approach. Future work will primarily focus on optimizing the speculative execution pipeline and further improving filesystem tracking in real-world settings.

\bibliographystyle{unsrt}
\bibliography{bibliography}

\appendix
\section{Appendix}

\subsection{Implementation Details}
\label{sec:impl_details}
As \fullname is designed to operate as a honeypot, comprehensive data capture is integrated throughout the execution pipeline. A primary objective is to support threat intelligence and forensic analysis of attacker methodologies. Consequently, the system employs verbose logging, recording all interactions, state changes, and internal module outputs for offline analysis.
The generation module relies on the \texttt{GPT-4.1-mini} model. This selection represents a balance between output quality, computational efficiency, and cost, allowing for extensive evaluation under realistic resource limitations. It is also motivated by prior work \cite{guan2024honeyllm,Honey_llama}, which demonstrates that the GPT family is particularly well suited for shell simulation tasks.
The speculative execution engine utilizes a two-stage prediction pipeline to anticipate likely future commands. First, a lightweight $n$-gram model estimates the probabilities of command sequences based on prior interaction history. Following empirical testing, we use a 3-gram architecture in which each token represents a distinct shell command. Second, a lightweight LLM predictor (GPT-5.4-mini) refines predictions by generating likely command options and arguments. Predicted commands are executed in parallel, allowing large, concurrent state explorations. 
The Smart Router leverages container-based isolation to provide a sandboxed execution environment for complex commands. The Memory Manager leverages the GPT-5.4-mini model to identify state-changing operations and update virtual file system state. The virtual file system is maintained through a JSON-based tree structure tracking file content, directories, and permissions. Finally, the adversarial attack detector uses GPT-5.4-mini to analyze input-output pairs and detect subversion attempts.

\paragraph{Code base} Additional information, including the dataset, specific prompt used, can be retrieved directly from our codebase: \repo

\subsection{Additional Experimental Results}
\label{app:additional_results}

\paragraph{Post-processor.}
The Post-processor defensive filtering does not harm performance on benign and on noisy commands, with a false positive rate below $0.01$ computed on the combination of the correctness and robustness tasks. Table~\ref{tab:detector_performance} provides the detailed performance results for the Post-processor filter.

\begin{table}[t]
\centering
\setlength{\tabcolsep}{5pt}
\caption{Classification performance of the Post-processor on correctness + robustness tasks.}
\label{tab:detector_performance}

\scriptsize
\begin{tabular}{cccc}
\toprule
\textbf{Precision} & \textbf{Recall} & \textbf{FPR} & \textbf{Balanced Accuracy} \\
\midrule
0.8947 & 0.9444 & 0.009 & 0.9677 \\
\bottomrule
\end{tabular}
\end{table}

\paragraph{Speculative Execution Engine.}
For the speculative execution module, we employ an $n$-gram-based predictor due to its fast inference capabilities and minimal computational overhead. We performed model selection by evaluating varying $n$-gram orders and Top-$k$ candidate thresholds. As reported in Table~\ref{tab:ngram_topk_selection}, a trigram model ($n=3$) yields the optimal performance. Specifically, configuring the predictor with $k=5$ provides the most effective trade-off between predictive accuracy and computational cost, achieving a validation accuracy of 0.645. On the unseen test set, this configuration maintains a robust hit rate of 0.658, indicating that the system correctly anticipates the subsequent user action in approximately 65.8\% of interactions. Increasing the threshold beyond $k=5$ introduces unnecessary computational overhead without yielding statistically significant accuracy gains. We stress that the predictor was trained on a dataset that is entirely disjoint from the user study dataset.

\begin{table}[t]
\centering
\setlength{\tabcolsep}{5pt}
\scriptsize
\caption{Validation accuracy across different $n$-gram orders and Top-$k$ selections. The best performance is achieved using a trigram ($n=3$) configuration.}
\label{tab:ngram_topk_selection}
\begin{tabular}{lccccc}
\toprule
& \multicolumn{5}{c}{\textbf{Top-$k$ Accuracy}} \\
\cmidrule(lr){2-6}
\textbf{$n$-gram} & \textbf{$k=1$} & \textbf{$k=2$} & \textbf{$k=3$} & \textbf{$k=5$} & \textbf{$k=10$} \\
\midrule
Unigram ($n=1$) & 0.190 & 0.344 & 0.375 & 0.413 & 0.450 \\ 
Bigram ($n=2$)  & 0.312 & 0.480 & 0.556 & 0.630 & 0.730 \\
Trigram ($n=3$) & \textbf{0.361} & \textbf{0.507} & \textbf{0.579} & \textbf{0.645} & \textbf{0.750} \\
4-gram ($n=4$)  & 0.344 & 0.475 & 0.534 & 0.585 & 0.680 \\
5-gram ($n=5$)  & 0.318 & 0.441 & 0.485 & 0.525 & 0.610 \\
6-gram ($n=6$)  & 0.294 & 0.415 & 0.451 & 0.487 & 0.560 \\
7-gram ($n=7$)  & 0.276 & 0.396 & 0.429 & 0.465 & 0.530 \\
8-gram ($n=8$)  & 0.261 & 0.386 & 0.419 & 0.454 & 0.510 \\
\bottomrule
\end{tabular}
\end{table}

\end{document}